\documentclass{ptephy}%%%%where ptephy is the template name

\preprintnumber{XXXX-XXXX} %%% Insert preprint number here

\usepackage{braket}

\usepackage[]{natbib}
\begin{document}

\title{Deformation and $\alpha$ clustering in excited states of $^{42}$Ca}

\author{\name{Yasutaka~Taniguchi (谷口\ 億宇)}{1,2}\thanks{Present address: Department of Medical and Basic Science, Nihon Institute of Medical Science, Moroyama-machi, Iruma-gun, Saitama 350-0435, Japan.}}
%%%%%%%%%%% The \name command should be used as \name{Insert author name here}{Insert affiliation number here}
%%%%% Please use \thanks for contributed author details

%%%%%%%%%%% The \affil command should be used as \affil{Insert affiliation number here}{Insert author address here}
\address{
\affil{1}{Center for Computational Sciences, University of Tsukuba, Tsukuba, Ibaraki 305-8577, Japan}
\affil{2}{RIKEN Nishina Center for Accelerator-Based Science, Wako, Saitama 351-0198, Japan}
}

\begin{abstract}%
Coexistence of various low-lying deformed states in $^{42}$Ca
and $\alpha$--$^{38}$Ar correlations in those deformed states have
been investigated using deformed-basis antisymmetrized molecular
dynamics. Wave functions of the low-lying states are obtained via
parity and angular momentum projections and the generator coordinate
method (GCM). Basis wave functions of the GCM calculation are
obtained via energy variations with constraints on the quadrupole
deformation parameter $\beta$ and the distance between $\alpha$ and
$^{38}$Ar clusters. The rotational band built on the $J^\pi = 0_2^+$
(1.84 MeV) state as well as the $J^\pi  = 0_3^+$ (3.30
MeV) state are both reproduced. The coexistence of two additional
$K^\pi = 0^+$ rotational bands is predicted; one band is shown to be
built on the $J^\pi = 0_3^+$ state. Members of the ground-state band
and the rotational band built on the $J^\pi = 0_3^+$ state contain
$\alpha$--$^{38}$Ar cluster structure components.
\end{abstract}

\subjectindex{xxxx, xxx}

\maketitle

\section{Introduction}

Drastic structural changes initiated by low excitation energies are
a significant characteristic of nuclear systems, and the coexistence
of deformed states and cluster structures is a typical phenomenon.
In the mass number region $A \sim 40$, low-lying normal-deformed
(ND) and superdeformed (SD) bands with many-particle-many-hole
(mp-mh) configurations have been confirmed experimentally in
$^{36,38,40}$Ar\cite{PhysRevC.63.061301,PhysRevC.65.034305,Ideguchi201018},
$^{40}$Ca\cite{PhysRevLett.87.222501},
$^{42}$Ca\cite{springerlink:10.1007/BF01412107,NSR2003LA04}, and
$^{44}$Ti\cite{PhysRevC.61.064314}. The SD band in $^{36}$Ar, ND and
SD bands in $^{40}$Ca, and SD band in $^{44}$Ti are considered to
have configurations of $4p8h$, $4p4h$, $8p8h$, and $8p4h$,
respectively, relative to the $sd$-shell double-closed structure.
Coupling of the cluster structure components in deformed states such
as the $\alpha$-cluster structure in the ND band of
$^{40}$Ca\cite{PTPS.132.7,PTPS.132.73,PTPS.132.103,taniguchi:044317}
and the ground-state band in
$^{44}$Ti\cite{PTPS.132.7,PTPS.132.73,Kimura200658} has also be
investigated.

In $^{42}$Ca, deformed states with mp-mh configurations and
clustering behavior have been observed experimentally, and the
rotational band built on the $J^\pi = 0_2^+$ (1.84 MeV) state
($K^\pi = 0_2^+$ band) has been
observed\cite{springerlink:10.1007/BF01412107,NSR2003LA04}. This
rotational band has a large moment of inertia that is similar to the ones of the SD bands in $^{36}$Ar and
$^{40}$Ca, scaled by $A^{5/3}$\cite{NSR2003LA04}, which is proportional to square of quadrupole deformation parameter $\beta$ in the liquid-drop model. In contrast to the small level
spacings, the in-band E2 transition strengths are rather weak and
are of the same order as those of the ND band in $^{40}$Ca. With
regard to $\alpha$-cluster structures, strong population to the
$J^\pi = 0_1^+$ and $0_3^+$ (3.30 MeV) states has been observed in
$\alpha$-transfer reactions to $^{38}$Ar, and the ratios of the
cross sections of $\alpha$ and $2n$ transfer reactions suggest that
the $J^\pi = 0_2^+$ and $0_3^+$ states have configurations of $6p4h$
and $4p2h$, respectively\cite{Fortune1978208}. Theoretically, the
$\alpha + ^{38}$Ar orthogonality condition model (OCM) describes $4p2h$
states with $\alpha$--$^{38}$Ar cluster structures, but a rotational
band with a $6p4h$ configuration is not obtained in low-lying
states\cite{PhysRevC.51.586}. To understand the structures in
$^{42}$Ca, various deformation with mp-mh configurations and
clustering should be taken into account, but such a study has never
been performed. The structures of low-lying states in $^{42}$Ca have
not yet been clarified.

This paper aims to clarify the structures of excited deformed bands in
positive-parity states of $^{42}$Ca by focusing on the coexistence
of rotational bands with mp-mh configurations. The
$\alpha$--$^{38}$Ar cluster correlations in low-lying deformed
states are also discussed.
To discuss coexistence and mixing of deformed and cluster structures, the generator coordinate method (GCM) are used.

This paper is organized as follows;
In Sec.~\ref{sec:framework}, the framework of this study is explained briefly.
In Sec.~\ref{sec:energy_variation}, results of energy variation to obtain GCM basis are shown.
In Sec.~\ref{sec:coexistence}, coexistence of various deformed states and their structures are discussed.
In Sec.~\ref{sec:E2_ph}, relations of E2 transition strengths and particle-hole configurations are discussed.
Finally, conclusions are given in Sec.~\ref{sec:conclusions}.

\section{Framework}
\label{sec:framework}

In this section, the framework of the study is explained briefly.
Details of the framework are provided in Refs.~\cite{PTP.93.115,PhysRevC.69.044319,PTP.112.475}.

\subsection{Wave function}

The wave functions in low-lying states are obtained by using the parity
and angular momentum projection (AMP) and the GCM with deformed-basis antisymmetrized molecular dynamics (AMD) wave functions. A deformed-basis AMD wave function
$\ket{\mathrm{\Phi}}$ is the Slater determinant of Gaussian wave
packets that can deform triaxially such that
\begin{eqnarray}
 &\ket{\mathrm{\Phi}} = \hat{\cal A}\ket{\varphi_1, \varphi_2, \cdots, \varphi_A},&\\
 &\ket{\varphi_i} = \ket{\phi_i} \otimes \ket{\chi_i} \otimes \ket{\tau_i},& \\
 &\langle \mathbf{r} | \phi_i \rangle = \pi^{-3/4} (\det {\mathsf K})^{1/2} \exp\left[ - \frac{1}{2} ({\mathsf K} \mathbf{r} - \mathbf{Z}_i)^2\right],& \\
 &\ket{\chi_i} = \chi^\uparrow_i \ket{\uparrow} + \chi^\downarrow_i \ket{\downarrow},& \\
 &\ket{\tau_i} = \ket{\pi}\ \mathrm{or}\ \ket{\nu},&
\end{eqnarray}
where $\hat{\cal A}$ denotes the antisymmetrization operator, and
$\ket{\varphi_i}$ denotes a single-particle wave function. The
$\ket{\phi_i}$, $\ket{\chi_i}$, and $\ket{\tau_i}$ denote the
spatial, spin, and isospin components, respectively, of each
single-particle wave function $\ket{\varphi_i}$. The real $3 \times
3$ matrix $\mathsf{K}$ denotes width of the Gaussian
single-particle wave functions that can deform triaxially, which is
common to all nucleons. The $\mathbf{Z}_i = (Z_{ix}, Z_{iy}, Z_{iz})$ are complex parameters to denote the centroid of
each single-particle wave function in phase space. The complex
parameters $\chi_i^\uparrow$ and $\chi_i^\downarrow$ denote the spin
directions. Axial symmetry are not assumed.

\subsection{Energy variation}

Basis wave functions of the GCM are obtained via the energy
variation with a constraint potential $V_\mathrm{cnst}$ after projection onto positive-parity states,
\begin{eqnarray}
 &\delta \left( \frac{\braket{\Phi^+ | \hat{H} | \Phi^+}}{\braket{\Phi^+ | \Phi^+}} + V_{\mathrm{cnst}} \right) = 0,&\\
 &\ket{\Phi^+} = \frac{1 + \hat{P}_r}{2} \ket{\Phi},&
\end{eqnarray}
where $\hat{H}$ is Hamiltonian, and $\hat{P}_r$ denotes the parity operator.
Variational parameters are
$\mathsf{K}$, $\mathbf{Z}_i$, and $\chi_i^{\uparrow,\downarrow}$ ($i= 1, ..., A$). The isospin component of each single-particle wave
function is fixed as a proton ($\pi$) or a neutron ($\nu$). The
Gogny D1S force is used as the effective interaction. 

To obtain deformed and cluster structure wave functions, two types of constraints $V_{\mathrm{cnst}}$ are used: the quadrupole deformation
parameter $\beta$ of the total system and the distance $d$ between
$\alpha$ and $^{38}$Ar clusters,
\begin{equation}
 V_\mathrm{cnst} =
  \left\{
   \begin{array}{l}
    v_\beta (\beta - \beta_0)^2\\
    v_d (d - d_0)^2
   \end{array}
  \right.
  .
\end{equation}
Here $\beta$ is the matter quadrupole deformation parameter.
The distance $d$ between $\alpha$ and $^{38}$Ar clusters are defined as distance between centers of mass of $\alpha$ and $^{38}$Ar clusters,
\begin{eqnarray}
 d &=& |\mathbf{R}_{\alpha\mbox{-}^{38}\mathrm{Ar}}|,\\
 R_{\alpha\mbox{-}^{38}\mathrm{Ar} \sigma} &=& \frac{1}{4} \sum_{i \in \alpha} \frac{\mathrm{Re} Z_{i\sigma}}{\sqrt{\nu_\sigma}} - \frac{1}{38} \sum_{i \in ^{38}\mathrm{Ar}} \frac{\mathrm{Re} Z_{i\sigma}}{\sqrt{\nu_\sigma}},
\end{eqnarray}
where $i \in \alpha$ and $^{38}$Ar mean that $i$th nucleon is contained in $\alpha$ and $^{38}$Ar clusters, respectively.
It should be noted that the $\sigma\ (=x, y, z)$ component of the spatial center of the single-particle wave function $\ket{\varphi_i}$ is $\frac{\mathrm{Re} Z_{i\sigma}}{\sqrt{\nu_\sigma}}$.
Details of the constraint of intercluster distance are provided in Ref.~\cite{PTP.112.475}.
When sufficiently large values are chosen for $v_\beta$ and $v_d$, the resultant values $\beta$ and $d$ become $\beta_0$ and $d_0$, respectively.

% For the energy variation with $\beta$,
%harmonic oscillator (HO) quanta for the protons and neutrons,
%$N_\pi$ and $N_\nu$, respectively, relative to the lowest allowed
%state are also constrained.
%The $N_\tau\ (\tau = \pi \mbox{ or } \nu)$ are defined as
%\begin{equation}
% N_\tau = \Braket{\frac{1}{2} (\mathsf{K} \hat{\mathbf{r}})^2 + \frac{1}{2} \left(\mathsf{K}^{-\mathrm{T}} \hat{\mathbf{k}}\right)^2} - \frac{3}{2} A - N_{0\tau},
%\end{equation}
%where $(N_{0\pi}, N_{0\nu}) = (, )$, which are HO quanta of lowest allowed states in $^{42}$Ca.

\subsection{Generator coordinate method}

  After performing the constraint energy variation for $\ket{\Phi^+}$, we
  superpose the optimized wave functions by employing the quadrupole
  deformation parameter $\beta$ and the distances $d$ between $\alpha$ and $^{38}$Ar clusters,
 \begin{equation}
    \Ket{\Phi^{J+}_M}
    = 
    \sum_K \hat{P}_{MK}^{J^\pi}
    \left(
    \sum_i f_{iK}^\beta \Ket{\Phi^\beta_i} 
    + \sum_i f_{iK}^d
    \Ket{\Phi^d_i}
    \right), \label{eq:HW} 
 \end{equation} 
where $\hat{P}_{MK}^{J^\pi}$ is the parity and total angular momentum
  projection operator, and $\Ket{\Phi^\beta_i}$ and $\Ket{\Phi^d_i}$ are optimized wave functions with $\beta$  and $d$ constraints for the constrained  values $\beta = \beta_0^{(i)}$ and $d = d_0^{(i)}$
  respectively.  
  The integrals over the three Euler angles in the total angular momentum projection operator $\hat{P}_{MK}^J$ are evaluated by numerical integration.  
  The numbers of sampling points in numerical integration are 23, 27 and 23 for $\alpha$, $\beta$ and $\gamma$, respectively. 
  Here the body-fixed $x$-, $y$- and $z$-axis are chosen as $\langle x^2 \rangle \leq \langle y^2 \rangle \leq \langle z^2 \rangle$ for $\gamma < 30^\circ$ wave functions and  $\langle x^2 \rangle \geq \langle y^2 \rangle \geq \langle z^2 \rangle$ for $\gamma > 30^\circ$ ones in the case of $\beta$-constrained wave functions. 
  In the case of $d$-constrained wave functions, the $z$-axis is chosen as the vector which connects the $\alpha$ and $^{38}$Ar clusters. 
  The coefficients $f_{iK}^\beta$ and $f_{iK}^d$ are
  determined by the Hill-Wheeler equation,  
  \begin{equation}
    \delta \left( \Braket{\Phi^{J+}_M | \hat{H} | \Phi^{J+}_M} - \epsilon \Braket{\Phi^{J+}_M | \Phi^{J+}_M}\right) = 0. 
  \end{equation}
  Then we get the energy spectra and the corresponding wave functions that expressed by the superposition of the optimum wave functions, $\{ \ket{\Phi^\beta_i} \}$ and $\{ \ket{\Phi^d_i} \}$.

%\section{Results}

\section{GCM basis obtained by energy variation}
\label{sec:energy_variation}

\begin{figure}[tbp]
 \begin{center}
  \includegraphics[width=0.75\textwidth]{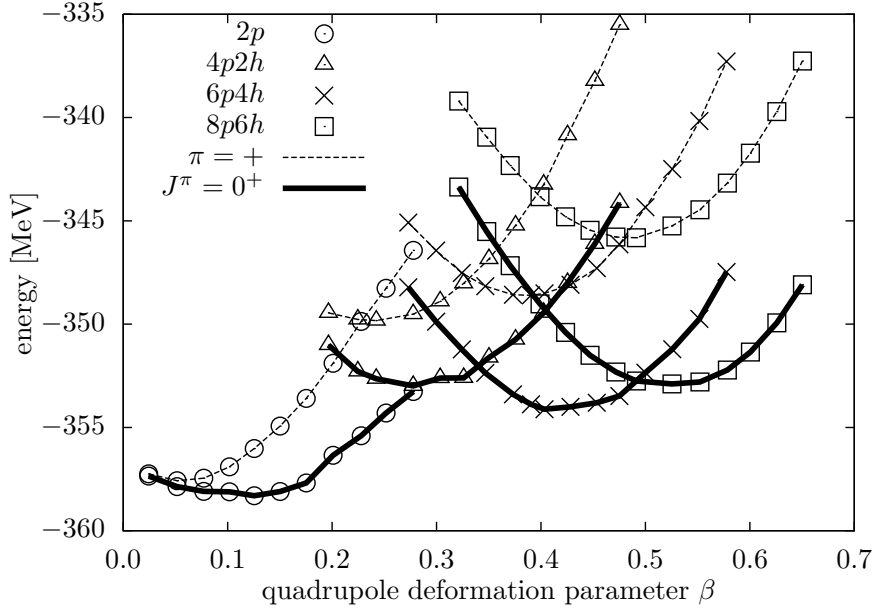}
  \caption{
  The energy curves as functions of the quadrupole deformation parameter $\beta$ for positive-parity (dashed lines) and $J^\pi = 0^+$ (solid lines) states.
 Circles, triangles, crosses, and squares indicate the $2p$, $4p2h$, $6p4h$, and $8p6h$ configurations, respectively (see text).
  }
  \label{42Ca_beta_energy}
 \end{center}
\end{figure}

\begin{figure}[tbp]
 \begin{center}
  \begin{tabular}{cc}
   \includegraphics[width=0.375\textwidth]{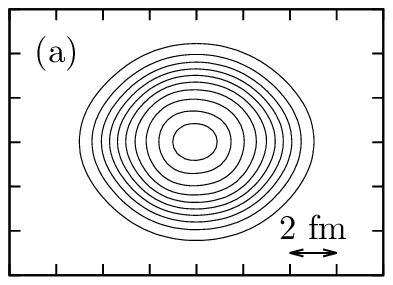} &
   \includegraphics[width=0.375\textwidth]{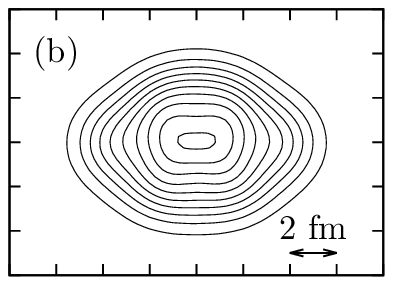} \\
   \includegraphics[width=0.375\textwidth]{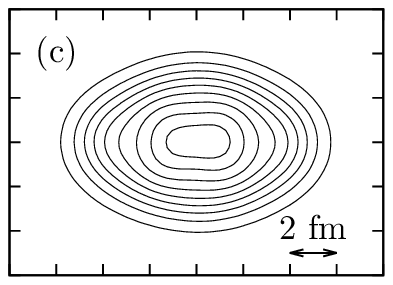} &
   \includegraphics[width=0.375\textwidth]{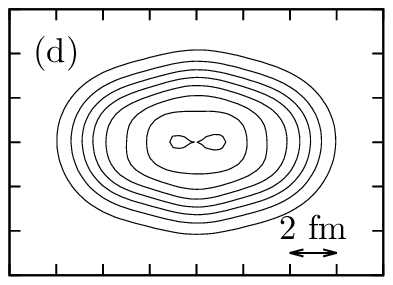} \\
   \includegraphics[width=0.375\textwidth]{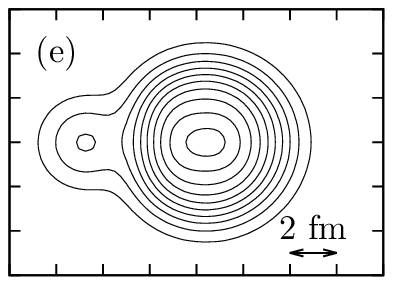} &
   \includegraphics[width=0.375\textwidth]{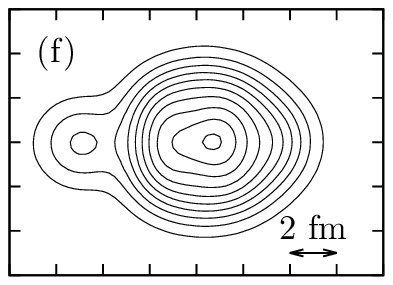} \\
  \end{tabular}
 \end{center} 
\caption{
 Density density distributions of (a) $2p$ ($\beta = 0.13$), (b) $4p2h$ ($\beta = 0.28$), (c) $6p4h$ ($\beta = 0.43$) and (d) $8p6h$ ($\beta = 0.53$) wave functions obtained with quadrupole deformation parameter $\beta$ constrained, and (e) A- and (f) B-type wave functions obtained with intercluster distance constraint ($d = $ 5.0 fm).
 }
 \label{fig:density}
\end{figure}

Figure~\ref{42Ca_beta_energy} shows the energy curves as functions of $\beta$, which are obtained by the energy variations with the constraint on $\beta$.
Harmonic-oscillator (HO) quanta of obtained wave functions for protons and neutrons, $N_\pi$ and $N_\nu$, respectively, are $(N_\pi, N_\nu) = (0, 0)$, (2, 0), (2, 2), and (4, 2) on and close to the $\beta$-energy surface relative to the lowest allowed state.
The $N_\tau$ ($\tau = \pi$ or $\nu$) are defined as
\begin{equation}
  N_\tau
   =
 \Braket{ \sum_{i \in \tau} \left[ \frac{1}{2} (\mathsf{K} \hat{\mathbf{r}}_i)^2 + \frac{1}{2} \left(\mathsf{K}^{-\mathrm{T}} \hat{\mathbf{k}}_i\right)^2 \right]} - \frac{3}{2} n_\tau
 -
 N_{0\tau},
\end{equation}
where $n_\pi$ and $n_\nu$ denote proton and neutron numbers, respectively, and $N_{0\pi}$ and $N_{0\nu}$ denote HO quanta of the lowest allowed states for protons and neutrons, respectively.
The (0, 0), (2, 0), (2, 2), and (4, 2) configurations correspond to
configurations of $[(pf)^2]_\nu$, $[(sd)^{-2}(pf)^2]_\pi [(pf)^2]_\nu$, $[(sd)^{-2}(pf)^2]_\pi [(sd)^{-2}(pf)^4]_\nu$, and
$[(sd)^{-4}(pf)^4]_\pi [(sd)^{-2}(pf)^4]_\nu$, respectively, which
in total are the $2p$, $4p2h$, $6p4h$, and $8p6h$ configurations,
respectively. The $4p2h$, $6p4h$, and $8p6h$ states have local
minima at $\beta \sim 0.3,$ 0.4, and 0.5, respectively. Deformations
of the protons and neutrons are similar across the whole $\beta$
region.
The $4p2h$, $6p4h$, and $8p6h$ states form triaxially
deformed structures. 
Density distribution of the wave functions at local minima with $2p$, $4p2h$, $6p4h$ and $8p4h$ configurations are shown in Figs.~\ref{fig:density} (a)-(d), respectively, which do not have significant neck structures.

Through the AMP, largely deformed states gain
higher binding energies, and the energy of the local-minimum state
for the $6p4h$ configuration projected onto the $J^\pi = 0^+$ state
becomes lower than that for $4p2h$ configuration. The order of the
local-minimum energies are the reverse of that before the AMP.
%Some wave functions with the $[(pf)^2]_\nu$ configuration contain much magnetized components, in which the time-reversal symmetry is broken.
%They have low energies projected onto $J^\pi = 2^+$ or $4^+$ states, and $2^+$ energies are lower than $0^+$ energies in $\beta \lesssim 0.05 $.
%Magnetized components contain no $J = 0$ components, and wave functions that contain much magnetized components are optimized mainly for $J > 0$ components by the energy variation.
%In small deformation region, magnetization can occur easily because level spacings of $0f_{7/2}$ orbits are small.
%$[(sd)^{-2}(pf)^2]_\pi [(pf)^2]_\nu$ configuration also have similar tendency.
%Unphysically low energies of magnetized components may be caused by the zero-range density-dependent term in the Gogny D1S force.
%It is because the zero-range density-dependent term makes strong repulsive force only for $(S, T) = (1, 0)$ channels.

\begin{figure}[tbp]
 \begin{center}
  \includegraphics[width=0.75\textwidth]{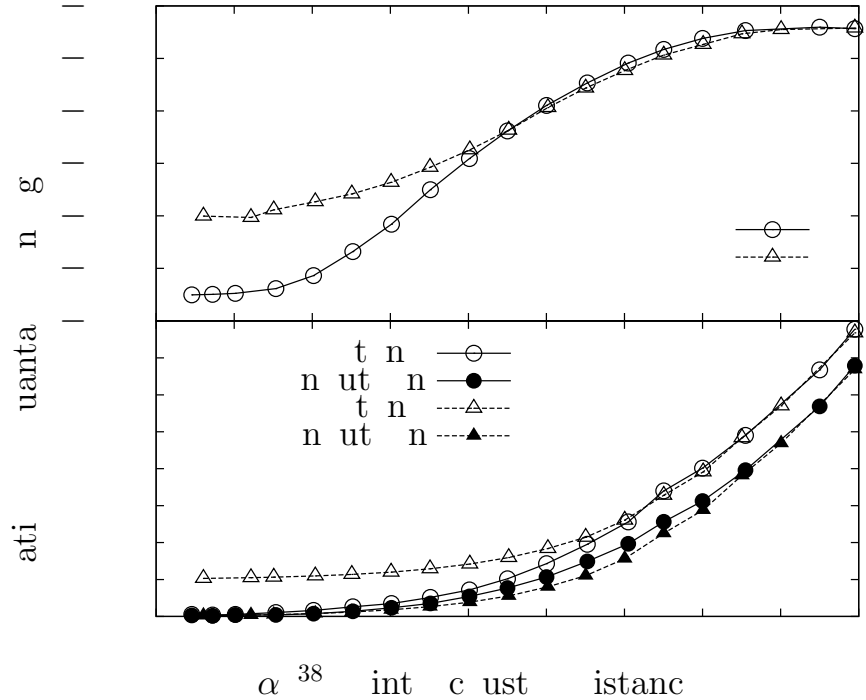}
  \caption{
  (upper) Solid and dashed curves show energies of $\alpha$--$^{38}$Ar cluster structures for A and B types (see text), respectively, as functions of intercluster distance.
  (lower) Solid and dashed curves show harmonic oscillator quanta for A and B types (see text), respectively, relative to the lowest-allowed state in $^{42}$Ca as functions of intercluster distance.
  Open and closed symbols are for protons and neutrons, respectively.
  }
  \label{A38Ar_energy}
 \end{center}
\end{figure}

Upper panel of Fig.~\ref{A38Ar_energy} shows energy curves of $\alpha$--$^{38}$Ar cluster structures as functions of intercluster distance between $\alpha$ and $^{38}$Ar clusters obtained by energy variations with the $\alpha$--$^{38}$Ar intercluster-distance constraint.
In the
calculations, two types, labeled A and B types, of $\alpha$--$^{38}$Ar cluster structure
wave functions are obtained that differ in the orientation of the
$^{38}$Ar clusters.
An $^{38}$Ar cluster has two proton-hole configuration relative to the $sd$-shell double-closed structure.
Proton holes of $^{38}$Ar clusters in A- and B-type wave functions are in parallel and orthogonal directions to an $\alpha$ cluster, respectively.
In small intercluster distance region, the A type has similar energies as the minimum energy on the $\beta$-energy surface.
But the B type is still excited, and the energies are similar to those at the local minimum of the $4p2h$ configuration of $\beta$-energy curves.
Lower panel of Fig.~\ref{A38Ar_energy} shows HO quanta for protons and neutrons, respectively, relative to the lowest-allowed state in $^{42}$Ca.
At small intercluster distance, the A type goes to the lowest-allowed state, whereas proton part of the B type are $2\hbar \omega$ excited, which is the $4p2h$ configuration.
It is because protons on the direction of an $\alpha$ cluster are occupied in $sd$-shell in the B type $^{38}$Ar clusters.
Owing to the Pauli principle, two protons are in the $pf$-shell even in small $\alpha$--$^{38}$Ar intercluster distance.
Density distributions of A- and B-type wave functions with $d = 5.0$ fm are shown in Figs.~\ref{fig:density} (e) and (f), respectively.
They have two spatially localized subsystems corresponding $\alpha$ and $^{38}$Ar clusters, which show $\alpha$--$^{38}$Ar cluster structures.
Shapes of $^{38}$Ar clusters are distorted due to intercluster interaction though ground state of $^{38}$Ar is almost spherical.

%\begin{figure}[tbp]
% \begin{center}
%  \includegraphics[width=0.5\textwidth]{42Ca_A38Ar_surface_HO.eps}
%  \caption{
%  Solid and dashed curves show harmonic oscillator quanta for A and B types (see text), respectively, relative to the lowest-allowed state of $^{42}$Ca as functions of intercluster distance.
%  Open and closed symbols are for protons and neutrons, respectively.
%  }
%  \label{A38Ar_HO}
% \end{center}
%\end{figure}

%\vspace*{1ex}

%\begin{figure}[tbp]
% \begin{center}
%  \includegraphics[width=0.5\textwidth]{42Ca_A38Ar_surface_energy.eps}
%  \caption{}
% \end{center}
%\end{figure}

%---$\alpha$-$^{38}$Ar---

%\vspace*{1ex}

\section{Coexistence of various rotational bands}
\label{sec:coexistence}

\subsection{Level scheme}

\begin{figure}[tbp]
 \begin{center}
  \includegraphics[width=0.75\textwidth]{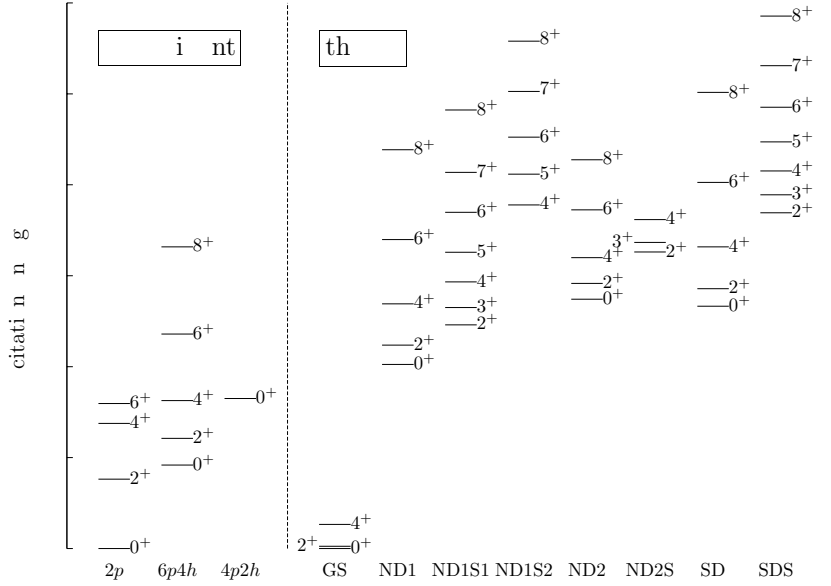}
  \caption{
  The experimental and theoretical level schemes in $^{42}$Ca.
  }
  \label{42Ca_level_scheme}
 \end{center}
\end{figure}

%\begin{table}[tbp]
% \caption{
% Squared overlaps of band-head states with space spaned by wave functions that have $2p$, $4p2h$, $6p4h$ and $8p6h$ configurations in the GS, ND1, ND2 and SD bands.
% }
% \label{tab:ph_config}
% \begin{ruledtabular}
%  \begin{tabular}{ccccc}
%   state & $2p$ & $4p2h$ & $6p4h$ & $8p6h$ \\
%   \hline
%   $0^+_{\mathrm{GS}}$  & 0.9 & 0. & 0. & 0. \\
%   $0^+_{\mathrm{ND1}}$ & 0. & 0. & 0.9 & 0. \\
%   $0^+_{\mathrm{ND2}}$ & 0. & 0.9 & 0. & 0. \\
%   $0^+_{\mathrm{SD}}$  & 0. & 0. & 0. & 0.9 \\
%  \end{tabular}
% \end{ruledtabular}
%\end{table}

Figure~\ref{42Ca_level_scheme} shows the level scheme of the
positive-parity states in $^{42}$Ca up to the $J^\pi = 8^+$ states
obtained via the AMP and the GCM. The GCM bases are
deformed-structure wave functions with configurations of $2p$,
$4p2h$, $6p4h$, and $8p6h$ obtained via energy variations with the
$\beta$ constraint and the $\alpha$--$^{38}$Ar
cluster structure wave functions obtained via energy variations
with the $\alpha$--$^{38}$Ar intercluster distance constraint to a
maximum of 9.0 fm.
Wave functions that contain more than 0.5 \% of the $J^+ K$ components $\Braket{\hat{P}^{J+}_{KK}}$ are adopted to GCM basis for each $JK$ to avoid numerical errors in the AMP.
Convergence of the GCM calculation was confirmed
by a comparison with a restricted set of basis wave functions (the
energies of states listed in Fig.~\ref{42Ca_level_scheme} change by less than 0.3 MeV when the
number of basis wave functions is halved). Three $K^\pi = 0^+$
rotational bands coexist in the excited states, labeled as ND1, ND2,
and SD.

\begin{figure}[tbp]
\begin{center}
  \includegraphics[width=0.75\textwidth]{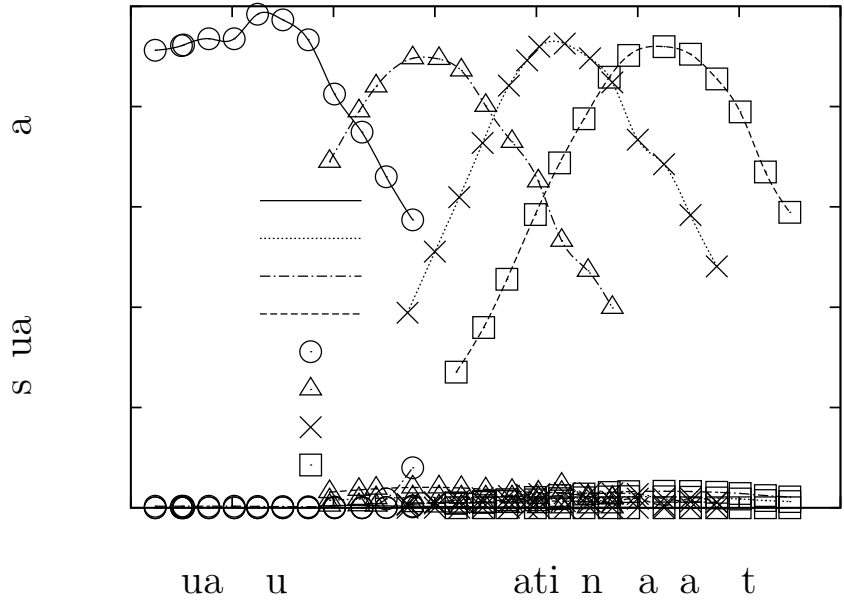}
 \caption{
 Squared overlaps of $J^\pi = 0^+$ states in the GS (solid), ND1 (dotted), ND2 (dot-dashed) and SD bands (dashed), and $J^\pi = 0^+$ components of wave functions obtained by energy variation with the $\beta$ constraint.
 Circle, triangle, cross and square symbols are for $2p$, $4p2h$, $6p4h$ and $8p6h$ wave functions, respectively.
 }
\end{center} 
\label{fig:beta_SO}
\end{figure}

Figure~\ref{fig:beta_SO} shows squared overlaps of $J^\pi = 0^+$ states and $J^\pi = 0^+$ components of wave functions obtained by energy variation with the $\beta$ constraint for the GS, ND1, ND2 and SD bands.
 The dominant components of the ND1, ND2 and SD states have $6p4h$, $4p2h$,
and $8p6h$ configurations, respectively, and the quadrupole
deformation parameters of their dominant components are $\beta =
0.43$, 0.28, and 0.53, respectively. The ground-state (GS) band has
a $2p$ configuration. The theoretical level spacings of the GS band
are underestimated although the GS band is considered to have a
simple $[(f_{7/2})^2]_\nu$ structure.
%The present framework does not assume a time-reversal symmetry and the $2^+$ and $4^+$ states contain low-energy magnetized components, which leads to the underestimation.
This underestimation does not affect qualitative properties of ND1, ND2 and SD bands because particle-hole configurations of their dominant components are much different from those of the GS band.

The present results suggest existence of side bands of the ND1, ND2, and SD bands due to triaxial deformation. The ND1S1 and ND1S2 are side bands
of the ND1 band whose dominant $K$ components are $|K| = 2$ and 4,
respectively. The ND2S and SDS bands are side bands of the ND2 and
the SD bands, respectively, with dominant components of $|K| = 2$.
The ND2 and SD bands, and the side bands of
the ND1, ND2, and SD bands, are theoretical predictions; candidate
states have not been observed yet.

\subsection{$\alpha$-$^{38}$Ar cluster correlations}

\begin{figure}[tbp]
 \begin{center}
  \includegraphics[width=0.75\textwidth]{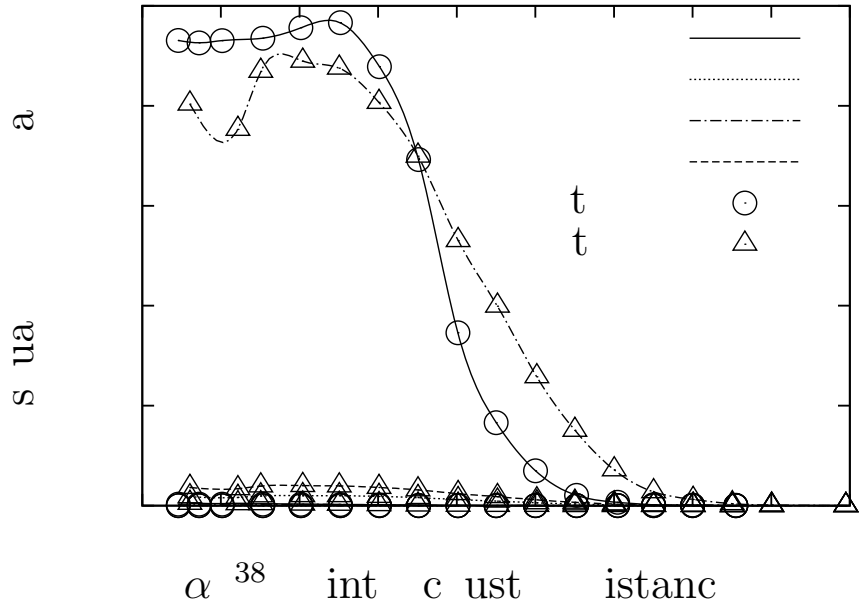}
  \caption{
  Squared overlaps of the $J^\pi = 0_\mathrm{GS}^+$ (solid), $0_\mathrm{ND1}^+$ (dotted), $0_\mathrm{ND2}^+$ (dot-dashed), and $0_\mathrm{SD}^+$ (dashed) states and the $\alpha$--$^{38}$Ar cluster structure wave functions as functions of the distance between $\alpha$ and $^{38}$Ar clusters.
  Open and closed symbols are for A- and B-type wave functions, respectively.
  }
  \label{A38Ar_amp}
 \end{center}
\end{figure}

To analyze the $\alpha$--$^{38}$Ar cluster structure correlations in
the low-lying rotational bands, squared overlaps of the band head
states and $\alpha$--$^{38}$Ar cluster structure components were
calculated for GS, ND1, ND2 and SD bands, as shown in Fig.~\ref{A38Ar_amp}.
The $J^\pi = 0_\mathrm{GS}^+$ and
$0_\mathrm{ND2}^+$ states have large amount of A- and B-type $\alpha$--$^{38}$Ar
cluster structure components, respectively, at large intercluster distances as well
as at small distances.
Particle-hole configuration of the A- and B-type $\alpha$--$^{38}$Ar cluster structure wave functions are equal to the the dominant particle-hole configurations of the $J^\pi = 0_\mathrm{GS}^+$ and $0_\mathrm{ND2}^+$ states, respectively, at small intercluster distance region, which shows particle-hole configuration of cluster wave functions at small intercluster distance are important for coupling to deformed states.
 The $J^\pi = 0_\mathrm{ND1}^+$ and $0_\mathrm{SD}^+$ states have small amount of $\alpha$--$^{38}$Ar cluster structure components for any intercluster distance.
Distributions of squared overlaps are similar up to high-spin state in each band.

\subsection{E2 transition strengths}

 \begin{table}
\begin{center}
   \caption{
  Theoretical and experimental $B(\mathrm{E2})$ values in Weisskopf
  unit $B_\mathrm{W.u.} = 8.67~\mathrm{e}^2\mathrm{fm}^4$ ($I_i$ and $I_f$ indicate initial and final states, respectively).
  Experimental values are taken from Refs.~\cite{springerlink:10.1007/BF01412107,Cameron2004293}.
  }
  \label{BE2}
   \begin{tabular}{cccc}
    \hline
           & $I_i$              & $I_f$              & $B(\mathrm{E2})$  \\
    \hline
    theory & $2^+_\mathrm{ND1}$ & $0^+_\mathrm{ND1}$ & 28.55 \\
           & $4^+_\mathrm{ND1}$ & $2^+_\mathrm{ND1}$ & 33.12 \\
           & $6^+_\mathrm{ND1}$ & $4^+_\mathrm{ND1}$ & 38.45 \\
 %           & $8^+_\mathrm{ND1}$ & $6^+_\mathrm{ND1}$ & 22.08 \\
    \cline{2-4}
           & $2^+_\mathrm{ND2}$ & $0^+_\mathrm{ND2}$ & 29.02 \\
           & $4^+_\mathrm{ND2}$ & $2^+_\mathrm{ND2}$ & 24.70 \\
           & $6^+_\mathrm{ND2}$ & $4^+_\mathrm{ND2}$ & 24.61 \\
 %           & $8^+_\mathrm{ND2}$ & $6^+_\mathrm{ND2}$ & 15.46 \\
    \cline{2-4}
           & $2^+_\mathrm{SD}$  & $0^+_\mathrm{SD}$  & 82.12  \\
           & $4^+_\mathrm{SD}$  & $2^+_\mathrm{SD}$  & 107.99 \\
           & $6^+_\mathrm{SD}$  & $4^+_\mathrm{SD}$  & 130.18 \\
 %           & $8^+_\mathrm{SD}$  & $6^+_\mathrm{SD}$  & 127.77 \\
    \hline
    experiment             & $4_2^+$ & $2_2^+$ & $57 \pm 42$      \\
    ($K^\pi = 0_2^+$ band) & $6_2^+$ & $4_2^+$ & $50^{+35}_{-16}$ \\
    \hline
   \end{tabular}
\end{center} 
\end{table}

The $B(\mathrm{E2})$ values of the in-band transitions in the
theoretical ND1, ND2, and SD bands, and the experimental $K^\pi =
0_2^+$ band in Weisskopf units are listed in Table \ref{BE2}. 
The theoretical values for the ND1 band are within error of the experimental values for the $K^\pi = 0_2^+$ band.
The in-band $B(\mathrm{E2})$ values from higher-spin state of the ND2 band are smaller, which is caused by mixing of components other than $4p2h$ configuration in higher-spin states.
The $B(\mathrm{E2})$ values of the SD band are much larger than those of ND1 and ND2 bands.

\subsection{Band assignment}

The amount of $\alpha$--$^{38}$Ar cluster components
(Fig.~\ref{A38Ar_amp}), particle-hole configurations of dominant components and the in-band transition $B(\mathrm{E2})$
values (Tab.~\ref{BE2}) indicate that the ND1 band and the ND2 band head correspond to the experimental $K^\pi = 0_2^+$ band
and $J^\pi = 0_3^+$ state, respectively. The large amount of
$\alpha$--$^{38}$Ar cluster components in the $J^\pi =
0_\mathrm{ND2}^+$ state reveals that this %$J^\pi = 0_\mathrm{ND2}^+$
state corresponds to the experimental $J^\pi = 0_3^+$ state because
of strong populations to the $J^\pi = 0_3^+$ state by
$\alpha$-transfer reactions to $^{38}$Ar\cite{Fortune1978208}, which
are sensitive to $\alpha$--$^{38}$Ar cluster structure components.
The ND1 states have small amount of $\alpha$--$^{38}$Ar cluster
structure components and similar in-band $B(\mathrm{E2})$ values to
those of the experimental $K^\pi = 0_2^+$ band, which indicates that
the ND1 band corresponds to the experimental $K^\pi = 0_2^+$ band.
The particle-hole configurations of the ND1 ($6p4h$) and ND2
($4p2h$) bands are consistent with those of the $J^\pi = 0_2^+$ and
$0_3^+$ states, respectively, suggested by the results of an
$\alpha$-transfer experiment\cite{Fortune1978208}. The members of
the ND2 band apart from the band head and those of the SD band have
not been observed yet. The members of the ND2 band could possibly be
observed by a combination of $\alpha$-transfer reactions and
$\gamma$-spectroscopy experiments because the ND2 band contains
large amount of $\alpha$-$^{38}$Ar cluster structure components and
has large in-band $B(\mathrm{E2})$ values.

This full-microscopic model reveals the coexistence of three
low-lying rotational $K^\pi = 0^+$ bands with $6p4h$, $4p2h$, and
$8p6h$ configurations in $^{42}$Ca, but the $\alpha$ + $^{38}$Ar
OCM, which is a semi-microscopic model, produces only one $K^\pi =
0^+$ rotational band with a $4p2h$ configuration in the low-lying
states. Full-microscopic models treating clustering and various
deformations with mp-mh configurations are required to understand
the low-lying structures in $^{42}$Ca. A unified treatment of
clustering and deformation is important for studying nuclear
structures.

\section{Particle-hole configurations and E2 transitions}
\label{sec:E2_ph}

The in-band $B(\mathrm{E2})$ values, deformations, and particle-hole
configurations in the ND1 and ND2 bands found here indicate that the
in-band $B(\mathrm{E2})$ values are more sensitive to the proton particle-hole configurations
 than to deformations, which are not
necessarily related to the $B(\mathrm{E2})$ values. Indeed, for ND1
and ND2 bands with the same proton particle-hole configurations, $(sd)^{-2}
(pf)^2$, the calculated in-band $B(\mathrm{E2})$ values are similar
although their dominant components have different quadrupole
deformations of $\beta = 0.40$ and 0.28, respectively. This is
inconsistent with a simple collective model in which the
$B(\mathrm{E2})$ values are proportional to $\beta^2$. Experimental
$B(\mathrm{E2})$ values of the in-band transitions in the ND band in $^{40}$Ca,
whose proton particle-hole configurations are also
$(sd)^{-2}(pf)^2$, are similar to the theoretical $B(\mathrm{E2})$
values of the ND1 and ND2 bands in $^{42}$Ca. As the Nilsson orbits show,
particle-hole configurations are strongly related to deformation,
but a careful consideration both of the particle-hole configurations
and deformation are required for understanding the structures of
deformed states.

\section{Conclusions}
\label{sec:conclusions}

In conclusion, the structures of the deformed states in $^{42}$Ca
have been investigated using deformed-basis AMD and the GCM by
focusing on the coexistence of various rotational bands with mp-mh
configurations and $\alpha$--$^{38}$Ar clustering. In the excited
states, three $K^\pi = 0^+$ bands, ND1, ND2, and SD, are obtained,
which have dominant $6p4h$, $4p2h$, and $8p6h$ configurations,
respectively. The ND1 band corresponds to the experimental $K^\pi =
0_2^+$ band, while the ND2 and SD bands have not yet been observed.
The band head of the ND2 band corresponds to the experimental $J^\pi
= 0_3^+$ state. The $B(\mathrm{E2})$ values of the in-band
transitions of the ND1 band are consistent with experimental data.
The members of the GS and ND2 bands contain large amount of $\alpha$--$^{38}$Ar
cluster structure components, which is consistent with results that
show the $J^\pi = 0_1^+$ and $0_3^+$ states are strongly populated
by $^{38}$Ar($^6$Li, $d$) reactions. Particle-hole configurations of
the dominant components of the GS and ND2 bands are consistent with
the suggestions of the $\alpha$-transfer to $^{38}$Ar experiment.
E2 transitions are more sensitive to proton particle-hole configurations than to deformation. It is necessary to employ
full-microscopic calculations and consider both clustering and
various deformations with mp-mh configurations for understanding the
low-lying states in $^{42}$Ca.

\section*{Acknowledgment}

The author thanks Dr. Y.~Kanada-En'yo and
Dr. M.~Kimura for careful proofreading and
valuable comments. Thanks are also given to Prof. H.~Horiuchi, Prof. K.~Ikeda, Dr. E.~Ideguchi, and Dr. M.~Niikura for fruitful discussions. This work has been supported by JSPS KAKENHI Grant Number 25800124 and University of Tsukuba Research Infrastructure Support Program.
 Numerical calculations
were conducted on the T2K-Tsukuba at Center for Computational Sciences, University of Tsukuba and the RIKEN Integrated Cluster of Clusters (RICC).

%Insert the Acknowledgment text here.


\bibliographystyle{ptephy}
\bibliography{42Ca_v6}
%
% once the .bbl file has been generated then place the text in your article.

%\vfill\pagebreak

\end{document}